\documentclass{ws-procs961x669}            % default, citations in superscript
\begin{document}
\title{Towards detecting gravitational waves: a contribution by Richard Feynman}

\author{M. Di Mauro}

\address{Dipartimento di Matematica, Universit\`a di Salerno,\\
Fisciano, 84084, Italy\\
E-mail: madimauro@unisa.it}

\author{S. Esposito$^*$ and A. Naddeo$^{**}$}

\address{INFN, Sezione di Napoli, C. U. Monte S. Angelo, Via Cinthia,\\
Napoli, 80125, Italy\\
$^*$E-mail: sesposit@na.infn.it\\
$^{**}$E-mail: anaddeo@na.infn.it}

\begin{abstract}
An account of Richard Feynman's work on gravitational waves is
given. Feynman's involvement with this subject can be traced back
to 1957, when he attended the famous Chapel Hill conference on the
Role of Gravitation in Physics. At that conference, he presented
in particular the celebrated sticky bead argument, which was
devised to intuitively argue that gravitational waves must carry
energy, if they exist at all. While giving a simple argument in
favor of the existence of gravitational waves, Feynman's thought
experiment paved the way for their detection and stimulated
subsequent efforts in building a practical detecting device.
Feynman's contributions were systematically developed in a letter
to Victor Weisskopf, completed in February 1961, as well as in his
Caltech Lectures on Gravitation, delivered in 1962-63. There, a
detailed calculation of the power radiated as gravitational
radiation was performed, using both classical and quantum field
theoretical tools, leading to a derivation of the quadrupole
formula and its application to gravitational radiation by a binary
star system. A comparison between the attitudes of Feynman and of
the general relativity community to the problems of gravitational
wave physics is drawn as well.
\end{abstract}


\bodymatter

\section{Introduction}\label{aba:sec1}
The contributions of Richard P. Feynman to physics notoriously
touched all four fundamental interactions of Nature. His first
breakthrough occurred in the theory of electromagnetism, with his
version of covariant quantum electrodynamics that earned him the
1965 Nobel prize\cite{Feynman:1965jda}, which was developed by him
between the late 1940s and the early 1950s. A few years later,
Feynman's interest had focused on another fundamental interaction,
namely gravity. Feynman first mentioned this new interest of his
to Murray Gell-Mann, who much later recalled:
\begin{quote}
We first discussed it when I visited Caltech during the Christmas
vacation of 1954-55 [...] I found that he had made considerable
progress\cite{Gell-Mann}.
\end{quote}
In view of this statement,\footnote{This was confirmed by Bryce S.
DeWitt, who declared: ``Even by 1955 Feynman claimed to have spent
a great deal of effort on the problem of
gravitation''\cite{Letter}.} it is conceivable that Feynman's
interests were directed towards gravity some time before 1954,
probably shortly after completing his work on quantum
electrodynamics. While working also on other subjects, most
notably condensed matter physics and the theory of weak
interactions\cite{Mehra:1994dz}, Feynman nurtured his interest in
gravity at least until the second half of the 1960s. In these
fifteen years, he gave several contributions to the both classical
and quantum gravity, some of which would later be recognized as
pivotal for the field. The first public contribution to gravity
was given by Feynman at the Chapel Hill conference in
1957\cite{ChapelHill}, where he took active part to many of the
numerous discussion sessions. After that, records of his later
work are documented by several published and unpublished sources.
For detailed historical accounts of Feynman's contributions to
gravitational physics, see Refs. \citenum{LectGravPreface} and
\citenum{DiMauro:2021yza}.

In this paper, we focus on Feynman's work on gravitational waves.
As well-known, at Chapel Hill Feynman proposed a simple and
intuitive thought experiment in favor of the existence of
gravitational waves. This ``sticky bead'' argument presumably
contributed to inspiring the pioneering experimental efforts by
Joseph Weber. In fact, Feynman's qualitative argument was
supported and extended by detailed calculations, several results
of which were presented at Chapel Hill as well. While the original
calculations are presumably lost, Feynman included a detailed
description of an updated version of them in a letter he wrote to
Viktor F. Weisskopf in early 1961\cite{WeisskopfLetter}. These
updated computations were also included in the famous graduate
lectures on gravitation he delivered at Caltech in the academic
year 1962-63\cite{Feynman:1996kb}. They show that Feynman had
addressed most of the contemporary theoretical problems of the
subject, showing that gravitational waves carry energy, and that
they can be detected by suitable instruments; moreover, he
computed in detail the energy radiated by binary self-gravitating
systems. Feynman's comments clearly show that he regarded his
results as compelling (and probably somewhat straightforward as
well, since he did not publish them) solutions to the above
mentioned issues. Indeed, in his subsequent work on gravity,
Feynman did not deal with gravitational waves any more, focusing
instead on quantum gravity.

The paper is organized as follows.  In Sect. \ref{History} we
briefly sketch the history of gravitational waves prior to
Feynman's engagement with them, highlighting the theoretical
problems which were the concern of scientists in the 1950s and
1960s. In Sect. \ref{ChapelHill} we briefly introduce the
historical context in which the Chapel Hill conference took place
and we summarize Feynman's contributions to it. In Sect.
\ref{WeisskopfLetter} we examine the letter to Weisskopf, and the
parts of the Lectures on Gravitation which discuss gravitational
waves. We close this section with a summary of Feynman's solutions
to the theoretical issues in gravitational wave research. In Sect.
\ref{Comments} we give some comments on Feynman's work on
gravitational waves, in particular in connection with his general
views on gravity and on fundamental interactions in general. In
Sect. \ref{Developments} we outline some contemporary and later
developments, and in particular we describe the solutions to the
theoretical issues that are currently accepted by the scientific
community. Sect. \ref{conclusions} is devoted to conclusions.

\section{Brief history of gravitational waves up to the 1950s}\label{History}

The main open problem concerning gravitational waves, when Feynman
tackled them, was their very existence as physical
entities.\footnote{For details on the history of gravitational
waves, with a focus on the theoretical side, we refer to Ref.
\citenum{Kennefick:2007zz}.} In fact, general relativity enjoys
general covariance, which means that coordinate systems have no
intrinsic meaning. While very elegant and powerful, this symmetry
also poses great difficulties, which have kept haunting general
relativity practitioners since the earliest days of the theory.
Namely, it is not easy to distinguish between real physical
phenomena and mere artifacts due to a bad coordinate choice, hence
invariant characterizations are needed. One of the earliest
examples is the Schwarzschild
singularity\cite{Schwarzschild:1916uq}, which occurs in the
spherically symmetric solution found by Karl Schwarzschild. In
fact, the metric describing this solution in spherical coordinates
displays a singularity on a surface located at a given value of
the radial coordinate. This singularity is now well-known to be
apparent, and it is removed by choosing a different coordinate
system\cite{Kruskal:1959vx}, but it took more than forty years to
really understand what this meant. Two other, strictly linked
examples, are given by the definition of energy in general
relativity, and by gravitational waves themselves. When Albert
Einstein first showed in 1916\cite{Einstein:1916cc} that the
gravitational field equations he had written down the year
before\cite{Einstein:1915ca} predict in the weak field limit the
existence of wave-like excitations, and when in
1918\cite{Einstein:1918btx} he corrected a major mistake in his
1916 paper, finding his celebrated quadrupole formula describing
the leading-order energy emission through gravitational radiation,
he chose a particular coordinate system, in which his equations
looked formally analogous to the equations of electromagnetism.
Einstein immediately recognized that some of the wave solutions he
had found were spurious, and could be eliminated by a coordinate
change (they were flat space in curved coordinates), while the
remaining ones were not spurious, and carried energy according to
his formula.

\subsection{The question of the existence of gravitational waves}

A weak point in Einstein's derivation was spotted and corrected by
Sir Arthur S. Eddington\cite{Eddington:1922ds}, namely, the fact
that Einstein's choice of coordinates explicitly contained the
speed of light, hence the propagation speed of gravitational waves
may have been inadvertently put in by hand by the German
physicist. Eddington improved and extended Einstein's work, giving
a more rigorous proof of the fact that some of Einstein's solution
were not coordinate artifacts, they carried energy and they
propagated with the speed of light. Finally, Eddington rederived
the quadrupole formula, correcting a minor error by Einstein (an
overall factor of 2), and applied it to the computation of energy
radiated away by a rigid rotator. However, in complaining about
the lack of an invariant way of distinguishing physical from
spurious waves, Eddington referred to the latter in a very
suggestive way, stating that, being coordinate artifacts, they
propagate with the ``speed of thought'' (Ref. \citenum{Eddington},
p. 130). This generated a misunderstanding according to which
Eddington attributed this characteristic to gravitational waves in
general, thus questioning their very existence. Indeed, doubts
concerning the existence of gravitational waves did not stop
bothering the scientists working in general relativity for a long
time. Another problem was posed by the fact that, unlike
electromagnetism, gravitation is described by a highly nonlinear
theory, while gravitational waves had been found only in the
linearized approximation. The possibility thus existed that
gravitational waves were in fact only artifacts of the linearized
theory, meaning that there is no solution to the full
gravitational field equations describing gravitational fields
propagating as waves. This point of view was shared for some time
by Einstein himself, after in 1937 he had tried with Nathan Rosen
to find plane wave solutions to the full gravitational field
equations\cite{EinsteinRosen1937}. Indeed, the two physicists
found singularities in their solutions, and interpreted them as
obstructions to the existence of propagating solutions to the
equations.\footnote{Einstein mentioned this in a letter to Max
Born (Ref. \citenum{EinsteinBorn}, letter 71).} After a well-known
incident with a referee\cite{KennefickPTOD}, Einstein finally
realized that the singularities he had found were coordinate
artifacts, and that in fact he and Rosen had found an exact
solution to his field equation, which enjoyed cylindrical
symmetry. This was one of the first solutions to the full theory
describing gravitational waves.\footnote{Actually, Einstein and
Rosen had rediscovered a solution first found by Guido Beck in
1925\cite{Beck:1925mm}.} However, the question of the physical
effects ascribable to these waves was still unclear, since it
seemed that they did not carry energy, as still argued by Rosen
himself in 1955 (Ref. \citenum{Bern}, pp. 171-174), and anyway the
Einstein-Rosen solution was non-physical, since its source was not
localized. What was needed was an exact solution describing waves
radiating from a localized center, which was still unavailable at
the time. In any case, the fact that even Einstein himself had
doubted that gravitational waves could exist certainly contributed
to making them a neglected field of study for twenty more years.

A very important breakthrough occurred in 1956, when Felix A. E.
Pirani managed to give an invariant characterization of
gravitational radiation, in terms of spacetime curvature rather
than energy\cite{Pirani1,Pirani2}, giving in fact birth to the
modern popular description of gravitational waves as propagating
ripples of curvature. This work was performed in the context of
Pirani's investigations of the physical meaning of the Riemann
curvature tensor, based on the geodesic deviation
equation\cite{Synge:1934zza} and on the classification of
curvature tensors in terms of their principal null
directions\cite{Petrov}. In particular, using the geodesic
deviation equation, Pirani showed that the passage of a
gravitational wave should modify the proper distance between test
particles, as an effect of the variation of the curvature. Thus, a
measurable effect could be ascribed at least in principle to
gravitational waves.

\subsection{The question of energy loss}

The issue of existence was not the only one which was at stake in
gravitational wave physics in the 1950s. Even taking existence for
granted, indeed, there was a problem concerning gravitational
radiation by self-gravitating systems, such as binary stars. The
first applications by Einstein and Eddington involved rigid bodies
such as rotators, which can be considered to be kept together by
the electromagnetic interaction, while a bound state kept together
by gravity itself is beyond the reach of the linearized
approximation, which was the only context in which gravitational
energy radiation had been studied. In fact, many physicists even
doubted that gravitational waves were radiated at all by such a
system. And even if radiation did occur, it was unclear whether
the Einstein quadrupole formula, which had been obtained in the
linearized theory, correctly described the resulting energy loss.
The issue had been addressed for the first time in 1941 by Lev D.
Landau and Evgenij M. Lifshitz, in the first edition of their
textbook \cite{Landau}, where the binary system was treated by
using the post-Newtonian approximation, and energy loss was
computed by applying the quadrupole formula. However, there was no
consensus about the correctness of their calculation, and the
question was still open in the 1950s.

\subsection{Summary of the open problems}
From what we said above, it emerges that the situation concerning
gravitational wave research was far from clear in the 1950s, with
many basic theoretical problems and no hope of experimental
guidance. We may summarize the main issues at stake as follows:
\begin{arabiclist}[3]
\item The actual existence of gravitational waves as physical
entities, their speed of propagation and their energy conveying;
\item The existence of gravitational waves as solutions to the
full nonlinear gravitational field equations; \item The radiation
of gravitational waves by self-gravitating systems and the
correctness of the quadrupole formula in describing the energy
loss.
\end{arabiclist}

\section{The Chapel Hill conference}\label{ChapelHill}

As we saw, before the 1950s research on gravitational waves was
carried out by only a handful of pioneers. This lack of interest
was in fact not only the result of the doubts expressed by
Einstein and also attributed to Eddington. The whole field of
general relativity remained dormant for thirty years, from the
mid-1920s to the mid-1950s, after an initial burst of activity.
During this period, later baptized ``low-watermark'' phase
\cite{Eisenstaedt1,Eisenstaedt2,Eisenstaedt3}, general relativity
was considered by most physicists to be a rather esoteric and
heavily mathematical subject, with tenuous relations with the
physical world.

Around the mid-1950's, things began to change, and gravity slowly
regained a prominent place in physics. Pivotal to this change of
attitude, which was dubbed the ``renaissance'' of general
relativity \cite{Will:1989rr,Will,Blum:2020oel}, was the
organization of a series of international conferences entirely
devoted to the subject. The renaissance of general relativity was
characterized by the fact that physicists began studying the
theory in a less speculative fashion than before, by focusing on
its physical implications, and by developing dedicated, i.e.
intrinsically general relativistic methods. That was also a time
where researchers from other fields, most notably particle and
field theory, began to be interested on gravitation and to apply
their methods to it. One of them was of course Feynman, who
started working on gravity just in this period, and it was in one
of these conferences, namely the one (organized by DeWitt and his
wife C\'ecile DeWitt-Morette) at the University of North Carolina
at Chapel Hill in January 1957\cite{ChapelHill}, that he first
announced his results in the field, as we anticipated in the
introduction. In his many contributions to the discussions at
Chapel Hill, he clearly delineated his attitude towards gravity,
and presented a host of results he had already obtained. Besides
the arguments on favor of the existence of gravitational waves,
which we discuss in the following subsections, Feynman sketched a
proposal for a field-theoretical approach to classical gravity.
His view, shared by other particle physicists, was indeed that
gravity had to be regarded as a fundamental interaction like all
the others, and hence it had to be described by a field theory,
which would also encode quantum corrections to the classical
theory. Indeed, within a few years, he managed to derive the field
equations of general relativity (as summarized in Ref.
\citenum{Feynman:1996kb}), and he attacked the problem of quantum
gravity, from this perspective
\cite{LectGravPreface,DiMauro:2021yza}. Finally, the Chapel Hill
conference is notably one of the few occasions where Feynman
expressed his views about foundational quantum issues, which were
inspired by the problem of the quantization of gravity
\cite{LectGravPreface,ChapelHill,DiMauro:2021yza,Zeh}.

\subsection{The sticky bead argument}

Feynman presented his famous sticky bead argument in a session
devoted to discussing the necessity of gravity quantization (in
which he also proposed a thought experiment in order to argue in
favor of that). In fact, Feynman had arrived a day later at the
conference\footnote{An amusing and well-known recollection of his
arrival in North Carolina can be found in Ref.
\citenum{FeynmanJoking}.}, skipping the session devoted to
gravitational waves. However, further discussion of gravitational
waves in sessions devoted to quantum gravity was not off-topic,
since the existence of these waves is of course very important for
the quantization of the theory. Thus, in the middle of a
discussion about whether the gravitational field had to be
quantized even in the absence of gravitational waves, Feynman
proposed a simple physical argument in favor of their existence
(Ref. \citenum{ChapelHill}, p. 260 and pp. 279-281). He argued
that if gravitational waves exist, they must carry energy and
this, together with the existence of these waves as solutions of
the linearized theory, was enough for him to be confident in their
existence. He expressed this intuitive feeling with the words:
``My instincts are that if you can feel it, you can make it" (Ref.
\citenum{ChapelHill}, p. 260). Feynman's reasoning was detailed in
an expanded version of his remarks, included in the records of the
conference\cite{ChapelHill}:
\begin{quote}
I think it is easy to see that if gravitational waves can be
created they can carry energy and do work. Suppose we have a
transverse-transverse wave generated by impinging on two masses
close together. Let one mass $A$ carry a stick which runs past
touching the other $B$. I think I can show that the second in
accelerating up and down will rub the stick, and therefore by
friction make heat. I use coordinates physically natural to $A$,
that is so at $A$ there is flat space and no field [...]. Then
Pirani at an earlier section gave an equation for the motion of a
nearby particle. (Ref. \citenum{ChapelHill}, p. 279).
\end{quote}
The result by Pirani is of course that obtained in Refs.
\citenum{Pirani1} and \citenum{Pirani2} (which indeed had been
presented earlier at the conference, cf. Ref.
\citenum{ChapelHill}, p. 141), stating that the displacement
$\eta$ of the mass $B$ (measured from the origin $A$ of the
coordinate system) in the field of a gravitational wave satisfies
the following differential equation:\footnote{This follows from
the geodesic deviation equation (Ref. \citenum{ChapelHill}, p.
141; Refs \citenum{Pirani1} and \citenum{Pirani2}).}
\begin{equation}\label{PiraniEq}
\ddot{\eta}^a+R^{a}_{0b0}\eta^b=0, \qquad (a,b=1,2,3)
\end{equation}
$R$ being the Riemann curvature tensor at the point $A$. Feynman
continued:
\begin{quote}
the curvature tensor [...] does not vanish for the
transverse-transverse gravity wave but oscillates as the wave goes
by. So, $\eta$ on the RHS is sensibly constant, so the equation
says the particle vibrates up and down a little (with amplitude
proportional to how far it is from $A$ on the average, and to the
wave amplitude.) Hence it rubs the stick, and generates heat.
(Ref. \citenum{ChapelHill}, p. 279)
\end{quote}
The upshot is that the stick absorbs energy from the gravitational
wave, which means that the latter carries energy and is able to do
work.

\subsection{Feynman's theoretical detector}

\begin{figure}
\centering
\includegraphics[width=2.5in]{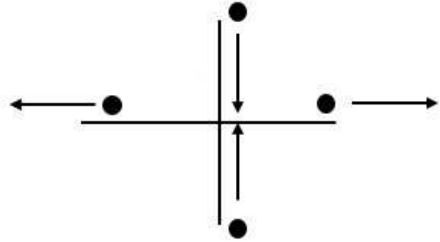}
\caption{Scheme of Feynman's gravitational wave detector.}
\label{aba:fig1}
\end{figure}

In connection with the device he considered, Feynman went on
considering a possible gravitational wave detector in which there
are four, rather than two test masses, which oscillate as shown in
Figure \ref{aba:fig1}.

This thought device was used to compute how much energy
gravitational waves would carry. The device was described by
Feynman as follows:
\begin{quote}
If I use 4 weights in a cross, the motions at a given phase are as
in the figure (Fig.\ref{aba:fig1}). Thus a quadrupole moment is
generated by the wave. Now the question is whether such a wave can
be generated in the first place. First since it is a solution of
the equations (approx.) it can probably be made. Second, when I
tried to analyze from the field equations just what happens if we
drive 4 masses in a quadrupole motion of masses like the figure
above would do - even including the stress-energy tensor of the
machinery which drives the weights, it was very hard to see how
one could avoid having a quadrupole source and generate waves.
Third my instinct is that a device which could draw energy out of
a wave acting on it, must if driven in the corresponding motion be
able to create waves of the same kind. [...] If a wave impinges on
our ``absorber'' and generates energy - another ``absorber''
placed in the wave behind the first must absorb less because of
the presence of the first, (otherwise by using enough absorbers we
could draw unlimited energy from the waves). That is, if energy is
absorbed the wave must get weaker. How is this accomplished?
Ordinarily through interference. To absorb, the absorber parts
must move, and \emph{in moving generate a wave} which interferes
with the original wave in the so-called forward scattering
direction, thus reducing the intensity for a subsequent absorber.
In view therefore of the detailed analysis showing that gravity
waves can generate heat (and therefore carry energy proportional
to $R^2$ with a coefficient which can be determined from the
forward scattering argument). I conclude also that these waves can
be generated and are in every respect real. (Ref.
\citenum{ChapelHill}, p. 280, emphasis in the original)
\end{quote}
In this quote, Feynman's point of view about the existence of
gravitational waves is evident: since they exist as solutions of
the field equations (albeit in the linearized approximation), they
carry energy, and they can be generated (in fact, the detector
generates secondary waves), then they must be a real phenomenon.

\subsection{Binary systems}

Within the discussion (Ref. \citenum{ChapelHill}, p. 260), Feynman
also quoted a result concerning the calculation of the energy
radiated by a self-gravitating two-body system in a circular
orbit. This result shows that he had addressed also this issue,
performing detailed calculations. The quoted result is:
\begin{equation}\label{doublestar}
\frac{\text{Energy radiated in one revolution}}{\text{Kinetic
energy content}}=\frac{16 \pi}{15} \frac{\sqrt{mM}}{m+M} \left(
\frac{u}{c} \right)^5,
\end{equation}
where $m$ and $M$ are the masses of the two bodies, and $u$ is
their relative velocity. In the case of the Earth-Sun system, this
formula describes as expected a tiny effect, leading to a huge
order of magnitude (about $10^{26}$ years) for the lifetime of the
motion of the Earth around the Sun. Despite the smallness of the
effect, this is a definite prediction. Feynman stated to have
performed a full analysis of this problem some time before (Ref.
\citenum{ChapelHill}, p. 280), but he only quoted the result,
without presenting his computations. Fortunately, Feynman reported
a version of them four years later, in the letter written to
Victor F. Weisskopf \cite{WeisskopfLetter}, to which we now turn.

\section{The letter to Weisskopf and the sixteenth lecture on
gravitation}\label{WeisskopfLetter}

As anticipated, a complete and systematic description of Feynman's
computations on gravitational waves can be found in a letter he
wrote to Weisskopf in February 1961 \cite{WeisskopfLetter}, to
answer a question that the Austrian physicist had asked him in the
fall of 1960, concerning the reality of gravitational radiation.
The letter was so detailed that it was subsequently included in
the material distributed to the students attending the Caltech
lectures in 1962-63. Indeed, the sixteenth lecture of that course
(the last one in the published version\cite{Feynman:1996kb}) dealt
with gravitational waves, and reproduced part of the computations
showed in the letter itself, with additional interesting comments.

The computations that Feynman reported in the letter and in the
lectures were surely related to those he had performed before the
Chapel Hill conference, but included several improvements that
Feynman had conceived meanwhile. Indeed, at the end of the letter,
he recalled the conference, claiming that
\begin{quote}
It was this entire argument used in reverse that I made at a
conference in North Carolina several years ago to convince people
that gravity waves must carry energy. [...] Only as I was writing
this letter to you did I find this simpler argument from the
Lagrangian (Ref. \citenum{WeisskopfLetter}, p. 14).
\end{quote}
What Feynman meant by ``argument used in reverse'' and by
``simpler argument from the Lagrangian'' will be clear by the end
of this section.

%\subsection{General considerations}
The most striking fact about the computations is that they are
performed adopting a quantum point of view, treating general
relativity as a quantum field theory, and then considering the
classical limit:
\begin{quote}
My view is quantum mechanical, but the classical limits are easily
derived. (Ref. \citenum{WeisskopfLetter}, p. 6)
\end{quote}
Only later the same computation is repeated from the usual,
classical point of view, in the framework of linearized gravity.
As we saw, in fact, in that period Feynman was struggling with the
quantization of gravity, building on the field theoretical
viewpoint that he had advocated at Chapel Hill. In the beginning
of 1961, he had recognized that loop diagrams in the theory
present major difficulties, but he was still far from a solution
(later in the same year he would report on his progress at the La
Jolla conference \cite{LaJolla}). However, he could safely study
classical gravity by applying his field theoretical methods to
tree diagrams, which are the only ones relevant to the classical
limit (while loop diagrams describe quantum corrections). This is
what he did in the letter,\footnote{A conceptually analogous
calculation was performed in 1936 by Matvei Bronstein
\cite{Bronstein}, who derived the quadrupole formula with the
correct factor of 2 by computing the emission coefficients of
transverse gravitons.} claiming that:
\begin{quote}
I am studying the problem of quantization of Einstein's General
Relativity. I am still working out the details of handling the
divergent integrals which arise in problems in which some virtual
momentum must be integrated over. But for cases of radiation,
without the radiation corrections, there is no difficulty. (Ref.
\citenum{WeisskopfLetter}, p.1)
\end{quote}
The ``radiation corrections'' Feynman was referring to are of
course what are usually called radiative corrections, i.e. higher
order corrections to the processes, which include the troublesome
loop diagrams. In fact, in the Lectures Feynman adopted this point
of view in studying many other problems in classical gravity:
\begin{quote}
All other problems in the theory of gravitation\footnote{Here
Feynman was referring to all problems, apart from that of a
spherical mass distribution and cosmology.} we shall attack first
by the quantum theory; in order to obtain the classical
consequences on macroscopic objects we shall take the classical
limits of our quantum answers. (Ref. \citenum{Feynman:1996kb},
p.163)
\end{quote}
The field theoretical methods that Feynman used are intrinsically
perturbative. Feynman explicitly discussed the validity of this
approximation, and of the linearized approximation he used in the
classical computation, stating that this is a natural and obvious
choice in view of the extreme weakness of the gravitational
interaction. He stated in fact that:
\begin{quote}
I won't be concerned with fields of arbitrary strength [...]. I am
surprised that I find people objecting to an expansion of this
kind, because it would be hard to find a situation where
numerically a perturbation series is more justified! (Ref.
\citenum{WeisskopfLetter}, p.1)
\end{quote}
which is confirmed in the Lectures:
\begin{quote}
There needs to be no apology for the use of perturbations, since
gravity is far weaker than other fields for which perturbation
theory seems to make extremely accurate predictions. (Ref.
\citenum{Feynman:1996kb}, p.207)
\end{quote}
From these quotations, it is evident that Feynman considered the
contemporary debate around the existence of gravitational waves in
the full nonlinear theory of general relativity, as opposed to the
linearized theory, to be pointless, since the
perturbative/linearized theory would furnish an excellent
approximation in most cases of interest.\footnote{He also
acknowledged the fact that the problem was at the moment of purely
academic interest, stating that ``the most ``practical'' thing to
do is to go to zero order and forget the whole problem
altogether'' (Ref. \citenum{WeisskopfLetter}, p.1). }

\subsection{The quantum computation}

To perform his computation, Feynman started from linearized
gravity, which from his point of view was the linear theory of a
massless charged spin-2 quantum field coupled to matter. For
simplicity, he considered scalar (spin-0) matter, and he computed
the probabilities for single graviton emission in several
processes involving matter particles, in the low energy limit. The
result is that the probability of emission of a graviton with
frequency $\omega$ in the solid angle $\mathrm{d}\Omega$ is (Ref.
\citenum{WeisskopfLetter}, p. 7; Ref. \citenum{Feynman:1996kb}, p.
214):
\begin{equation}\label{quantum}
P=a^2\frac{\mathrm{d}\Omega}{4\pi}\frac{\mathrm{d}\omega}{\omega}\frac{\lambda^2}{4\pi^2},
\end{equation}
where $a$ is a kinematical factor and $\lambda$ is the
gravitational coupling constant (which is proportional to the
square root of Newton's constant $G$).

Feynman knew that the probabilities of radiation of massless
quanta become universal in the low energy limit, i.e. the
probability of emission of a soft quantum in an interaction is
independent of the spin of the interacting quanta (hence the
choice of scalar matter implies no loss of generality), and most
importantly of the kind of interaction which is involved. This
happens because the low-energy limit is equivalent to the
long-wavelength limit, in which the physics is insensitive to
short-distance details. In particular, soft graviton emission
occurs, with the same probability, also when matter quanta
interact gravitationally. Since classical radiation is related to
soft quantum emission, the implication is that masses accelerating
under the reciprocal gravitational interactions must emit
gravitational radiation as any other accelerating mass:
\begin{quote}
Although this was worked out for spin zero particles, the result,
as $\omega\rightarrow 0$, is the same for all, spin 1/2, photons,
etc. It is correct for collisions irrespective of the kind of
force. For example, Schiff asked me if two masses under their
mutual gravitation can radiate. The answer is certainly yes. (Ref.
\citenum{WeisskopfLetter}, p.7)
\end{quote}
According to Feynman, these results proved that self-gravitating
system must radiate gravitational waves. This view was confirmed
in the Lectures:
\begin{quote}
As far as the radiation is concerned, the exact nature of the
over-all scattering process is not important. I emphasize the last
point because there are always some theorists who go about
mumbling some mystical reasons to claim that the radiation would
not occur if the scattering is gravitational - there is no basis
for these claims; as far as we are concerned, radiation of gravity
waves is as real as can be; the sun-earth rotation must be a
source of gravitational waves. (Ref. \citenum{Feynman:1996kb}, p.
215).
\end{quote}
Thus the issue of gravitational radiation by self-gravitating
systems was considered solved by Feynman. We emphasize that his
quantum point of view was pivotal to that conclusion, since his
reasoning relied on a well-known result of quantum field theory,
namely the universality of soft graviton emission. Once again,
Feynman showed some annoyance towards the physicists who continued
to have doubts, while the solution was so clear to him.

\subsection{The classical computation and the quadrupole formula}

In the second part of the letter, Feynman switched to a classical
computation, which was considered to be more appropriate for
macroscopic objects, such as celestial bodies:
\begin{quote}
for the motion of big objects such as planets and stars it may be
more consistent to work in the classical limit. (Ref.
\citenum{Feynman:1996kb}, p. 215)
\end{quote}
The classical computation was developed in close analogy with
electrodynamics, taking into account that in the case of
gravitation the source is a tensor $S_{\mu\nu}$ rather than a
vector $J_{\mu}$. The differential equation which has to be solved
to find classical gravitational waves is (in the notation of Ref.
\citenum{Feynman:1996kb}):
\begin{equation}\label{waves1}
\Box^2 \overline{h}_{\mu\nu}=\lambda S_{\mu\nu},
\end{equation}
where $h_{\mu\nu}$ is the field describing deviations of the
metric from the flat Minkowski metric and the bar operation is
defined as:
\begin{equation} \label{barOperation}
\overline{h}_{\mu\nu}=\frac{1}{2} \left( h_{\mu\nu} + h_{\nu\mu}
\right) - \frac{1}{2} \eta_{\mu\nu} {h^{\sigma}}_{\sigma}.
\end{equation}
$\Box^2$ is the usual, flat-space, d'Alembertian operator. As
well-known, the above equation corresponds to the linearized
Einstein equations in the de Donder gauge
$\partial^{\mu}\overline{h}_{\mu\nu}=0$. When all quantities are
periodic with a given frequency $\omega$, the solution at a point
$1$ located at a distance from the source much greater than the
linear dimensions of the source itself (which is the region where
$S_{\mu\nu}$ is expected to be large):
\begin{equation}\label{SolWaves1}
\overline{h}_{\mu\nu}(\vec r_1)=- \frac{\lambda}{4\pi r_1} e^{i
\omega r_1} \int d^3 \vec r_2 S_{\mu\nu}(\vec r_2) e^{-i \vec K
\cdot \vec r_2},
\end{equation}
with $|\vec r_1|\gg |\vec r_2|$ as stated. This approximation
holds for nearly all cases of astronomical interests, where
wavelengths are much longer than the system's dimensions, such as
for instance binary stars and the Earth-Sun system. Finally,
Feynman assumed that the motions of matter in the source were
slow, so that he could resort to the post-Newtonian approximation.
This is where the computations in the last section of Lecture 16
of the Caltech Lectures (Ref. \citenum{Feynman:1996kb}, pp.
218-220) stopped, while in the letter Feynman went on computing
the power radiated by the above waves in the quadrupole
approximation which, for a periodic motion of frequency $\omega$,
read (Ref. \citenum{WeisskopfLetter}, p.11):
\begin{equation}\label{PowerRad}
\frac{\text{mean energy radiated}}{\text{sec.}}=\frac{1}{5} G
\omega^6 \sum_{ij} \left| Q_{ij}^{'} \right|^2,
\end{equation}
where $Q_{ij}^{'}=Q_{ij}-\frac{1}{3} \delta_{ij} Q_{kk}$,
$Q_{ij}=\sum_{a} m_a R_i^a R_j^a$ is the quadrupole mass moment,
and the root mean square average is taken. The result
(\ref{PowerRad}) was then shown to agree with the quantum
mechanical probability (\ref{quantum}) computed in the first part
of the letter (after suitable averaging). To strengthen this
result, Feynman had considered earlier (Ref.
\citenum{WeisskopfLetter}, p. 10) an analogous comparison for the
case of electromagnetism, finding again perfect correspondence.

Feynman then applied his formulas to the case of a circularly
rotating double star (Ref. \citenum{WeisskopfLetter}, p. 12),
promptly recovering the result (\ref{doublestar}). In sum,
Feynman's classical computation was analogous to that performed
earlier by Landau and Lifshitz. However, his treatment was
simpler, since he managed by using a very ingenious method for
computing the energy density, which essentially exploits the fact
that for classical transverse plane waves the linearized gravity
Lagrangian reduces to that of the harmonic oscillator. This spared
him formal complications such as the cumbersome energy-momentum
pseudo-tensor, which was used by Landau and Lifschitz. This is in
fact the ``simpler argument from the Lagrangian'' which he
referred to in the end of the letter.

The above treatment was then supplemented with a thorough analysis
of the effects of a gravitational wave impinging on a device made
of two test particles placed on a rod with friction, in this way
providing a more complete description of the working principle of
the gravitational wave detector previously introduced at Chapel
Hill. A second absorber, made of four moving particles in a
quadrupole configuration, as in Figure \ref{aba:fig1}, was also
considered, and shown to be able to absorb energy from a
gravitational wave acting on it. The oscillating device was shown
to re-radiate waves with the same energy content. Apparently, at
Chapel Hill, Feynman had derived the expression for the radiated
energy (which was then applied to the binary system case) from the
detailed study of the quadrupolar detector, either because at that
time he did not know how to extract the definition of energy from
the gravitational Lagrangian, or because the existing methods were
considered to be too cumbersome by him. In the letter, instead,
Feynman found a simple way of taking the opposite route (from this
we can understand why Feynman claimed that the argument at Chapel
Hill had been ``used in reverse''), since the quadrupole formula
was derived in general from the linearized Einstein equations and
then applied to the detector case.

Thus, Feynman had shown that gravitational waves carry energy, and
he had computed how much, also in the case of a binary star,
satisfactorily addressing both the first and the third of the
above issues. At the end of Lecture 16 of Ref.
\citenum{Feynman:1996kb}, Feynman commented again on the energy
content of gravitational waves:
\begin{quote}
What is the power radiated by such a wave? There are a great many
people who worry needlessly at this question, because of a
perennial prejudice that gravitation is somehow mysterious and
different - they feel that it might be that gravity waves carry no
energy at all. We can definitely show that they can indeed heat up
a wall, so there is no question as to their energy content. The
situation is exactly analogous to electrodynamics (Ref.
\citenum{Feynman:1996kb}, p. 219).
\end{quote}
and in the letter, while recalling the Chapel Hill conference he
commented:
\begin{quote}
I was surprised to find a whole day at the conference devoted to
this question, and that ``experts'' were confused. That is what
comes from looking for conserved energy tensors, etc. instead of
asking  ``can the waves do work?'' (Ref.
\citenum{WeisskopfLetter}, p. 14),
\end{quote}
These two quotes again display Feynman's annoyance towards the
relativity community. The fact that relativists kept arguing about
these problems without reaching an agreement, and that they
privileged sophisticated formal tools to plain physical thinking,
definitely disturbed Feynman, who repeatedly expressed his
annoyance, and may have contributed to his loss of interest in the
subject.

\subsection{Feynman's answer to the three issues}

The detailed study of Feynman's work on gravitational waves
reveals what he thought the solutions to the issues listed in
Sect. \ref{History} should be, at least from a theoretical point
of view, pending the ultimate experimental verification. In
summary, these were:
\begin{arabiclist}[3]
\item Feynman's thought experiments and calculations clearly show
that gravitational waves exist and that they carry energy, while
the fact that they propagate with the speed of light is a
consequence of the fact that they emerged from massless quanta,
i.e. the gravitons; \item Since gravity is so weak, it makes
perfect sense to study gravitational waves using the linearized
approximation; \item Gravitationally-bound system emit
gravitational radiation as any other bound system, and the energy
loss is described by the quadrupole formula in the approximations
valid in most situations of interest.
\end{arabiclist}

\section{Comments on Feynman's attitude towards gravity and gravitational waves}\label{Comments}

As emphasized, Feynman's approach to gravitational waves, and to
gravitation in general, was much closer in spirit to particle
physics than to general relativity. In fact, Feynman preferred a
field theoretical approach over the geometrical one which had
become the standard one among relativists. In his opinion, indeed,
the latter approach divided gravity from other fundamental
interactions, preventing the investigation of the relation of
gravitation with the rest of physics, and also posing great
obstacles towards its quantization.

A second motivation which underlies Feynman's choice of using
quantum field theoretical methods for investigating classical
gravitational radiation and other problems in classical gravity,
was his belief that quantum mechanics underlies Nature at the most
fundamental level, with the forces we observe at the macroscopic
level emerging from quantum interactions in the classical limit.
This is most clearly expressed by him in some unpublished
lectures:
\begin{quote}
I shall call conservative forces, those forces which can be
deduced from quantum mechanics in the classical limit. As you
know, Q.M. is the underpinning of Nature... (Ref.
\citenum{FeynmanHughes2}, p. 35)
\end{quote}
Feynman recognized that the main problem with gravity was the fact
that general relativity had received only a handful of
experimental confirmations, while the subjects of gravitational
waves and quantum gravity were confined to pure theory. These
subjects were being investigated by the relativity community by
resorting to sophisticated mathematics, in the hope that rigor
could be a substitute for empirical evidence. Feynman had a
different point of view. To him, the best way to proceed was to
try to extract physical predictions from the theory \emph{as if}
these were easily accessible experimentally. This viewpoint was
strongly advocated by him at Chapel Hill in some critical
comments:
\begin{quote}
There exists, however, one serious difficulty, and that is the
lack of experiments. [...] so we have to take a viewpoint of how
to deal with problems where no experiments are available. There
are two choices. The first choice is that of mathematical rigor.
[...] The attempt at mathematical rigorous solutions without
guiding experiments is exactly the reason the subject is
difficult, not the equations. The second choice of action is to
``play games'' by intuition and drive on. Take the case of
gravitational radiation. [...] I think the best viewpoint is to
pretend that there are experiments and calculate. In this field
since we are not pushed by experiments we must be pulled by
imagination. (Ref. \citenum{ChapelHill}, pp. 271-2)
\end{quote}
and shortly after
\begin{quote}
The real challenge is not to find an elegant formalism, but to
solve a series of problems whose results could be checked. This is
a different point of view. Don't be so rigorous or you will not
succeed. (Ref. \citenum{ChapelHill}, p. 272)
\end{quote}
This is in fact the viewpoint that Feynman adopted in his
investigations of gravitational waves, but also in his work on
quantum gravity \cite{DiMauro:2021yza}.

\section{Contemporary and subsequent developments}\label{Developments}

When the Chapel Hill conference took place, the time was ripe for
the existence of gravitational waves to be accepted by most
physicists. Thus, the work of Pirani, and Feynman's sticky bead
argument, managed to convince people that gravitational waves
belong to the realm of physical phenomena, and suggested ways to
investigate their physical effects. In fact, an argument similar
to Feynman's was conceived independently by a leading relativist,
Hermann Bondi, who was also present at Chapel Hill. This is
referred to by several of Bondi's remarks at the conference, such
as the following one, made after Pirani's talk (Ref.
\citenum{ChapelHill}, p.142): ``Can one construct in this way an
absorber for gravitational energy by inserting a $d\eta/d\tau$
term, to learn what part of the Riemann tensor would be the energy
producing one, because it is that part that we want to isolate to
study gravitational waves?''. The absorber of gravitational energy
is just a more formal description of the ``stickiness'' of
Feynman's sticky-bead. Shortly after the conference, Bondi
published his variant of the sticky-bead device \cite{Bondi},
although unlike Feynman he did not succeed in relating the
intensity of the gravitational wave to the amount of energy
carried by it. Inspired by these discussions, another participant
to the conference, Joseph Weber, soon undertook his lifetime
effort to experimentally detect gravitational waves \cite{Weber1};
Weber built his famous resonant bar detector in 1966 \cite{Weber2}
and even announced the first experimental detection of
gravitational waves coming from the center of the Milky Way in
1969 \cite{Weber3}. While this finding was later disproved,
Weber's works triggered all the subsequent research on the
detection of gravitational waves, which culminated in the
monumental detections by the LIGO-Virgo collaboration in
2016\cite{LIGOScientific:2016aoc}. Among all the physical results
that this brought about, the confirmation came that gravitational
waves move with the speed of light \cite{LIGOScientific:2017zic}.

Thus, it may be said that the first issue, that of the existence
and physical effects of gravitational waves, was solved after
Chapel Hill, with agreement among the scientific community. The
same cannot be said of the other two issues. The relativity
community reached an agreement on the status of gravitational
waves in the full nonlinear theory only some years later, thanks
to the rigorous work performed in that framework by Bondi, Pirani,
Ivor Robinson and Andrzej Trautman, among others. In particular,
Robinson and Trautman found exact solutions to the equations of
general relativity describing waves radiated from bounded sources
\cite{Robinson:1960zzb}. Furthermore, in the sixties, Bondi,
Rainer W. Sachs, Ezra T. Newman and Roger Penrose
\cite{Bondi:1962px,Sachs:1962wk,Sachs:1964zza,Newman:1961qr,Penrose:1965am},
gave a satisfactory definition of the energy radiated at infinity
as gravitational radiation by an isolated gravitating system in an
asymptotically flat spacetime. Thus, the modern theoretical
characterization of gravitational waves had emerged by the late
1960s.

The last issue took still longer to be settled, since it crucially
involved the problem of motion in general relativity and the issue
of radiation reaction, whose foundations were still quite shaky
\cite{Ehlers:1976ji}. The doubts concerning whether binary systems
radiated gravitational energy were fought by crucial empirical
input, which came in the mid-1970s with the discovery of the first
binary neutron star\cite{Hulse:1974eb,HulseTaylor2}.\footnote{In
fact, so far, all direct and indirect evidence for gravitational
waves has come from binary systems, which hence are the only
confirmed sources.} This did not immediately settle the quadrupole
formula controversy yet \cite{Kennefick:2016mxm}, but it prompted
many efforts towards its resolution. The confirmation of came when
Thibault Damour\cite{Damour:1983tz} showed that there was
quantitative agreement between the emission rate computed from the
quadrupole formula and observations. In fact, it is only in the
final stages of the merging of compact objects that higher (mass
and current) momenta need to be included (see e.g. Ref.
\citenum{Blanchet:2013haa} and references therein).

\section{Conclusions}\label{conclusions}

In the above pages, we have described the work performed by
Richard Feynman in the physics of gravitational waves. We have
seen how his attitude towards physics, and towards fundamental
interactions, shaped this work, and how he managed to give sound
and substantially correct answers to the main theoretical problems
of the time. Feynman's physical intuition and sound calculations
convinced him early on of the existence of gravitational waves and
gravitational radiation. His work contributed to the settlement of
the issue, and played a role in triggering the first experimental
efforts towards their detection. Feynman did not share the
theoretical concerns of most relativists, who continued to argue
about other issues, until rigorous solutions and/or experimental
input were available, and considered those arguments pointless,
harshly criticizing them. This may have been contributed to his
loss of interest in the subject, starting from the late 1960s,
when he began to be involved to the only fundamental interaction
to whose understanding he had not contributed yet, that is, the
strong interaction\cite{Mehra:1994dz}. But that is another story.

\section*{Acknowledgments}

The authors would like to thank the staff of the Caltech archives
for providing them a copy of Feynman's letter to Weisskopf
\cite{WeisskopfLetter}.

%Non BiBTeX users can list down their references as:

\end{document}